\documentclass[useAMS,usenatbib]{mn2e}
\pdfoutput=1
\usepackage{enumitem}
\usepackage{natbib}
\usepackage{graphicx}
\usepackage{amssymb}
\usepackage{lineno}
\usepackage{amsmath}

\begin{document}
\newcommand{\MSun}{{M_\odot}}
\newcommand{\LSun}{{L_\odot}}
\newcommand{\Rstar}{{R_\star}}
\newcommand{\calE}{{\cal{E}}}
\newcommand{\calM}{{\cal{M}}}
\newcommand{\calV}{{\cal{V}}}
\newcommand{\calO}{{\cal{O}}}
\newcommand{\calH}{{\cal{H}}}
\newcommand{\calD}{{\cal{D}}}
\newcommand{\calB}{{\cal{B}}}
\newcommand{\calK}{{\cal{K}}}
\newcommand{\labeln}[1]{\label{#1}}
\newcommand{\Lsolar}{L$_{\odot}$}
\newcommand{\farcmin}{\hbox{$.\mkern-4mu^\prime$}}
\newcommand{\farcsec}{\hbox{$.\!\!^{\prime\prime}$}}
\newcommand{\kms}{\rm km\,s^{-1}}
\newcommand{\cc}{\rm cm^{-3}}
\newcommand{\Alfven}{$\rm Alfv\acute{e}n$}
\newcommand{\Vap}{V^\mathrm{P}_\mathrm{A}}
\newcommand{\Vat}{V^\mathrm{T}_\mathrm{A}}
\newcommand{\D}{\partial}
\newcommand{\DD}{\frac}
\newcommand{\TAW}{\tiny{\rm TAW}}
\newcommand{\mm }{\mathrm}
\newcommand{\Bp }{B_\mathrm{p}}
\newcommand{\Bpr }{B_\mathrm{r}}
\newcommand{\Bpz }{B_\mathrm{\theta}}
\newcommand{\Bt }{B_\mathrm{T}}
\newcommand{\Vp }{V_\mathrm{p}}
\newcommand{\Vpr }{V_\mathrm{r}}
\newcommand{\Vpz }{V_\mathrm{\theta}}
\newcommand{\Vt }{V_\mathrm{\varphi}}
\newcommand{\Ti }{T_\mathrm{i}}
\newcommand{\Te }{T_\mathrm{e}}
\newcommand{\rtr }{r_\mathrm{tr}}
\newcommand{\rbl }{r_\mathrm{BL}}
\newcommand{\rtrun }{r_\mathrm{trun}}
\newcommand{\thet }{\theta}
\newcommand{\thetd }{\theta_\mathrm{d}}
\newcommand{\thd }{\theta_d}
\newcommand{\thw }{\theta_W}
\newcommand{\beq}{\begin{equation}}
\newcommand{\eeq}{\end{equation}}
\newcommand{\ben}{\begin{enumerate}}
\newcommand{\een}{\end{enumerate}}
\newcommand{\bit}{\begin{itemize}}
\newcommand{\eit}{\end{itemize}}
\newcommand{\barr}{\begin{array}}
\newcommand{\earr}{\end{array}}
\newcommand{\bc}{\begin{center}}
\newcommand{\ec}{\end{center}}
\newcommand{\DroII}{\overline{\overline{\rm D}}}
\newcommand{\DroI}{{\overline{\rm D}}}
\newcommand{\eps}{\epsilon}
\newcommand{\veps}{\varepsilon}
\newcommand{\vepsdi}{{\cal E}^\mathrm{d}_\mathrm{i}}
\newcommand{\vepsde}{{\cal E}^\mathrm{d}_\mathrm{e}}
\newcommand{\lraS}{\longmapsto}
\newcommand{\lra}{\longrightarrow}
\newcommand{\LRA}{\Longrightarrow}
\newcommand{\Equival}{\Longleftrightarrow}
\newcommand{\DRA}{\Downarrow}
\newcommand{\LLRA}{\Longleftrightarrow}
\newcommand{\diver}{\mbox{\,div}}
\newcommand{\grad}{\mbox{\,grad}}
\newcommand{\cd}{\!\cdot\!}
\newcommand{\Msun}{{\,{\cal M}_{\odot}}}
\newcommand{\Mstar}{{\,{\cal M}_{\star}}}
\newcommand{\Mdot}{{\,\dot{\cal M}}}
\newcommand{\ds}{ds}
\newcommand{\dt}{dt}
\newcommand{\dx}{dx}
\newcommand{\dr}{dr}
\newcommand{\dth}{d\theta}
\newcommand{\dphi}{d\phi}

\newcommand{\pt}{\frac{\partial}{\partial t}}
\newcommand{\pk}{\frac{\partial}{\partial x^k}}
\newcommand{\pj}{\frac{\partial}{\partial x^j}}
\newcommand{\pmu}{\frac{\partial}{\partial x^\mu}}
\newcommand{\pr}{\frac{\partial}{\partial r}}
\newcommand{\pth}{\frac{\partial}{\partial \theta}}
\newcommand{\pR}{\frac{\partial}{\partial R}}
\newcommand{\pZ}{\frac{\partial}{\partial Z}}
\newcommand{\pphi}{\frac{\partial}{\partial \phi}}

\newcommand{\vadve}{v^k-\frac{1}{\alpha}\beta^k}
\newcommand{\vadv}{v_{Adv}^k}
\newcommand{\dv}{\sqrt{-g}}
\newcommand{\fdv}{\frac{1}{\dv}}
\newcommand{\dvr}{{\tilde{\rho}}^2\sin\theta}
\newcommand{\dvt}{{\tilde{\rho}}\sin\theta}
\newcommand{\dvrss}{r^2\sin\theta}
\newcommand{\dvtss}{r\sin\theta}
\newcommand{\dd}{\sqrt{\gamma}}
\newcommand{\ddw}{\tilde{\rho}^2\sin\theta}
\newcommand{\mbh}{M_{BH}}
\newcommand{\dualf}{\!\!\!\!\left.\right.^\ast\!\! F}
\newcommand{\cdt}{\frac{1}{\dv}\pt}
\newcommand{\cdr}{\frac{1}{\dv}\pr}
\newcommand{\cdth}{\frac{1}{\dv}\pth}
\newcommand{\cdk}{\frac{1}{\dv}\pk}
\newcommand{\cdj}{\frac{1}{\dv}\pj}
\newcommand{\rad}{\;r\! a\! d\;}
\newcommand{\half}{\frac{1}{2}}

\title[
Evidence for pulsars metamorphism]
{Evidence for pulsars metamorphism and their possible connection to black holes and dark matter in cosmology{}}
{}
\author[Hujeirat,  A.A.]
       {Hujeirat, A.A. \thanks{E-mail:AHujeirat@uni-hd.de} \\
\\
IWR, Universit\"at Heidelberg, 69120 Heidelberg, Germany \\
}
\date{Accepted  ...}

\pagerange{\pageref{firstpage}--\pageref{lastpage}} \pubyear{2002}

\maketitle

\begin{abstract}

Pulsars and neutron stars are generally more massive than the Sun, whereas
black holes have unlimited mass-spectrum, though the mass-gap
between 2 - 5$\MSun,$ which applies for both classes, is evident and
remains puzzling.

Based on the solution of the TOV equation modified to include a universal scalar field $\phi$ at the background of supranuclear densities, we claim that pulsars must be born with embryonic super-baryons (SBs), that form through the merger of individual neutrons at their centers. The cores of SBs are made of purely incompressible superconducting gluon-quark superfluids (henceforth SuSu-fluids). Such quantum fluids have a uniform supranuclear density and governed by the critical EOSs  for baryonic matter $P_b = \calE_b$ and for
$\phi-$induced dark energy $P_\phi= -\calE_\phi.$\\

   The incompressibility here ensures that particles communicate on the shortest possible timescale, superfluidity and superconductivity enforce SBs to spin-down promptly as dictated by the Onsager-Feynman equation, whereas their lowest energy state grants
   SBs lifetimes that are comparable to those of  protons.
   The extraordinary lo ng lifetimes suggests that conglomeration of SuSu-objects
   would evolve over several big bang events to possibly form dark matter halos that
   embed the galaxies in the observable universe.

   Having pulsars been converted into SuSu-objects, which is predicted to last for one Gyr or even shorter, then they become extraordinary compact and turn invisible.\\
   It turns out that recent observations on the quantum, stellar and cosmological scales   remarkably support the present scenario.  \\
\end{abstract}


\section{Superfluidity in pulsars}
Pulsars and NSs are considered to be made of superfluids governed by triangular lattice of quantized vortices as prescribed by the Onsager-Feynman
equation: $\oint \textbf{v}\cdot \textbf{d} l = \DD{2\pi \hbar}{ m}N. $  $\textbf{v},\textbf{d}l,~ \hbar, m$ here denote the velocity field, the vector of line-element, the reduced Planck constant and the mass of the superfluid particle pair, respectively.\\
Accordingly,  Crab pulsar should have approximately $N_n=  8.6\times10^{17}$
neutron and $N_p \approx 10^{30}$ proton-vortices (Fig. 1).
Let the evolution of the number density of vortex lines, $n_v$, obeys the following  advection-diffusion equation:
\beq
            \DD{\D n_v}{\D t} + \nabla\cdot n_v \textbf{u}_f = \nu_t  \triangle n_v,
\eeq
where $t,\, \textbf{u}_f,~\nu_t$ denote the transport velocity at the cylindrical radius $r=r_f$ and dissipative coefficient in the local frame
of reference, respectively. When $\nu_t=0$, then
the radial component of  $\textbf{u}_f $ in cylindrical coordinates reads: \\
$u^{max}_f\approx-(\dot{\Omega}/{\Omega})~r >0.$ In the case of the Crab;
this would imply that approximately $10^6$  neutron  vortices must be expulsed/annihilated each second,  and therefore the object should switch off  after $10^{6}$ or $10^{13}$ yr,
depending on the underlying mechanism of heat transport \cite[see][and the references therein]{Baym1995,Link2012}. However both scenarios are contrary to observation, as numerous NSs  has been found, which are older than $10^{6}$ yrs, though non of them is older than  $10^{9}$ yrs.
\begin{figure}
\centering {
\includegraphics*[angle=-0, width=7.15cm]{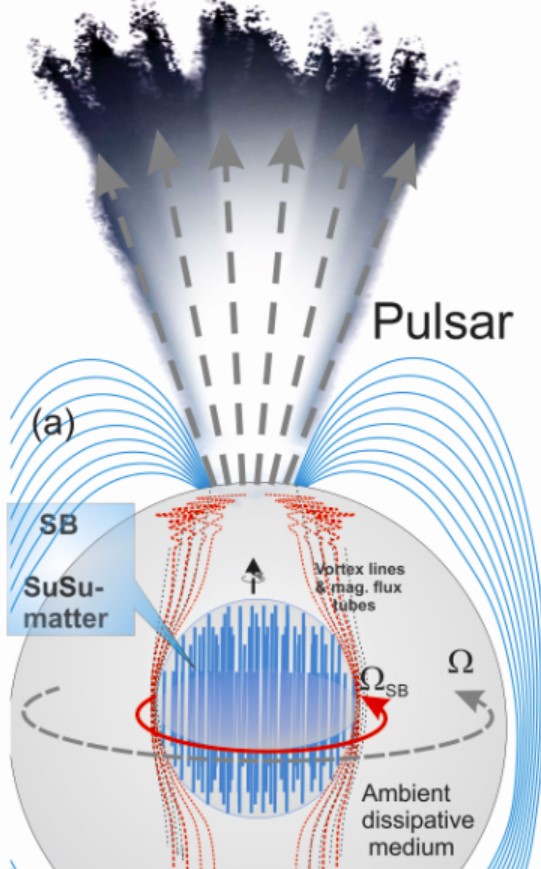}\\
}
\caption{\small  A magnetized pulsar is born with an embryonic rigid-body rotating super-baryon (SB) at its center, which is made of incompressible gluon-quark superfluid. By interacting with the ambient medium, the SB expels certain number of vortex lines that are absorbed by the surrounding dissipative medium, thereby causing to promptly spin up: the exact quantum signatures of the glitches observed in pulsars and young neutron star systems.   }  \label{NSVortices}
\end{figure}
On the other hand, recent numerical calculations of superfluids reveal generation of large amplitude Kelvin waves
that turn superfluids turbulent \cite[see][and the references therein]{Baranghi2008, Baggaley2014, Dix2014}.
It is therefore unlikely that trillions of Kilometer-long neutron and protons-vortices inside pulsars and NSs would behave differently.
 In this case,  $u_f$ should be replaced by  a mean turbulent  velocity $<u_f>^t $ with  $ u^{max}_f$ being an upper
 limit\footnote{The rotational energy associated with the outward-transported vortex lines from the central regions
are turbulently re-distributed in the outer shells and should not necessary suffer of a complete annihilation.}.
 As the number of vortex lines decreases with time due to emission of
 magnetic dipole radiation and therefore the separation between them increases non-linearly, it is reasonable to associate a time-dependent turbulent length scale
 $\ell_t(t),$ which covers the two limiting cases: $\ell_t(t=0) = \ell_0 \approx 10^{-3}\,$cm and  $\ell_t(t=\infty)=\ell_{\infty}= R_\star.$
 This yields the geometrical mean   $<\ell_t> = \sqrt{\ell_0 \ell_{\infty}}\approx \calO(10)\,cm.$
  Putting terms together and using $\nu_{tur}= <\ell_t><u_f>^t $  to describe the effective turbulent viscosity, we obtain an upper limit for the global diffusion time scale:
$\tau_{diff} = {R_{NS}^2}/{\nu_{tur}}= \calO(10^9)$ yr.   Similarly,  a comparable time scale for the
Ohmic diffusion in this turbulent medium can be constructed as well.
This is in line with observations, which reveal that most isolated luminous NSs known are younger than $10^9$ yr   \citep[see][and the references therein]{Espinoza2011}.\\
On the other hand, as stable degenerate NSs require the density gradient to be negative,  then the very central region  would be the first to be evacuated from vortex lines and all other removable energies that do not contribute significantly to the pressure.  Let $r_f$ be the radius of the central region. The star cools and loses energy, $r_f$  would creep outwards with an average velocity:  $\dot{r}_f\sim R_\star/\tau_{diff} \approx10^{-10}$ cm/s.
As I show in the next sections, the nuclear matter inside $r_f$  would undergo a  transition from compressible dissipative neutron fluid into an incompressible gluon-quark superfluid phase; the lowest possible energy state. Once $r_f = R_\star - \epsilon$ $(\epsilon \ll 1),$ then the object turns invisible.\\

   In analogy with normal massive luminous stars, we expect massive NSs to also switch-off  earlier than their less massive counterparts.
 The reason is that most models of EOSs for NSs predict a correlation of $M_{NS} \propto \alpha_s,$ where  $\alpha_s (\doteq \DD{R_S}{R_\star})$
 denotes the compactness parameter, $R_\star, R_S$ are star and the corresponding Schwarzschild  radii \citep{Haensel2007}. In the extremal case however, when $\alpha_s = 1,$ it is reasonable  to expect that there will be no turbulence to dissipate. In this case   the expression of for the global turbulent diffusion time scale should be modified as follows:
 $\tau_{diff} = {R_{NS}^2}/{\nu_{tur}}\rightarrow {(R_{NS} - R_S)^2  }/{\nu_{tur}}. $  In terms of NS-parameters, the turbulent viscosity reads:
 $\nu_{tur}= R_{NS}^2  ({\dot{\Omega}}/{\Omega})  \sqrt{({\ell_0 }/{R_{NS}})}. $  Putting terms together, we obtain
 $\tau_{diff} = (1- \alpha_s)^2 \sqrt{({R_{NS}}/{\ell_0 })}({\Omega}/{\dot{\Omega}}).$ For a Vela-type pulsar with $\alpha_s \approx 2/3,$
the effect of compactness would shorten $\tau_{diff}$ by almost one order of magnitude.

\section{The incompressibility of supranuclear dense fluids}
Modeling the internal structure of cold NSs while constraining their masses and radii  to observations, would
  require their central densities to inevitably be much higher than the nuclear density $-\rho_0$:
a density regime in which all EOSs become rather uncertain and mostly acausal \citep[see][and the references therein]{Hampel2012}. On the other hand, at very high densities, almost all EOSs converge to the stiffest EOS: $P_{local} = \calE$
\citep{Camenzind2007},where fluids become purely incompressible.

The corresponding chemical potential here reads:
\beq
              \mu =\DD{\D \calE_b}{\D n} = \DD{P_{local}+\calE_b}{n}= \DD{2\calE_b}{n},
\eeq
whose solution is: 
\[
               \calE_b = a_\infty~ n^2.
\]
Particles obeying this EOS communicate with the speed of the light. This implies that the number
density must be upper-bounded by $n_{cr},$ beyond which local thermodynamical quantities become constants, specifically $\mu = \calE_\phi = P_{local} = const.$  In this case calculating the degenerate
pressure from the local quantities becomes invalid as it yields a vanishing local pressure
$(P_{local}= n^2 \DD{\D}{\D n}(\calE_b/n) = 0).$\\
Even if the fluid were weakly-compressible only, $\nabla P_{local}$ would be too smooth to hold the object gainst its own self-gravity. This becomes even blatant at the center, where
the gradient of the pressure vanishes to meet the regularity condition.
  The usual adopted strategy to escape this pressure-deficiency is to enforce  an unfounded inward-increase of $\calE_b$ as $r\rightarrow 0,$ resulting  therefore in unreasonably large central densities.\\

 Alternatively, one could argue that the strong gravitational field would enforce fusing of neutrons at the center of NSs and form  a sea of gluon-quark fluid.  However, a phase transition from normal nuclear matter into a gluon-quark plasma (GQP) under classical conditions was found to be unlikely (Baym \& Chin 1976; Chapline \& Nauenberg 1977).
 The GQPs in these studies were assumed to be locally conformal, compressible and governed by a non-local and constant bag pressure.\\

However, the situation may differ if there is a mechanism at the background of supranuclear densities, that among others, is capable of injecting energy into the system.  This in turn would effectively increase the mass of the object, steepen the curvature of the embedding spacetime and compress the nuclear fluid up to the saturation level, beyond which it
becomes purely incompressible. The underlying conjecture here is that compression would mainly affect the distance between individual neutrons, $\ell_n,$ rather than the distance, $\ell_q,$ between the quarks enclosed in individual neutrons. \\

In fact there are at least two observational facts that in favor of this argument:
\begin{enumerate}
  \item Massless gluons mediate the strong force between quarks with the speed of light,
  whereas global compression of the central fluid-core proceeds almost in a quasi-stationary manner. Hence, as the quark contribution to baryon mass is negligibly small, compressibility of the gluon-quark plasma would violate the causality principle.\\

  \item The quark potential ${\cal{V}}_q$ inside baryons increases with the radius to attains  maximum at the boundaries - $r_b$. Quark confinement however requires that ${\cal{V}}_q(r_b) \approx 0.94$ GeV.  In this case the
       effective/reactive velocity at the boundary would read:
        $<V> = \Delta x/\Delta t =  \Delta x \,\Delta E /h \approx c,$ where we set $\Delta x = r_b= 10^{-13}$ cm. Consequently, the reaction time of GQPs inside neutrons to whatsoever external compressional effects is of order $10^{-24}$ sec: the shortest possible time scale in the entire system.\\
      Moreover, recalling that the lifetime of neutrons in free space is extremely
      short compared to that of protons, then compression of GQPs inside neutrons
      most likely would provoke a runaway decay into protons and  would give rise to neutrino-dominated electromagnetic eruptions. However, such events can be safely ruled out by observations.
  \item Recent RHIC and LHC experiments have revealed that the matter resulting from smashed protons behaves like fluids, whose constituents move collectively and relatively slowly. These observations are in line with the basic  properties of incompressible fluids rather than randomly moving particles in plasmas.\\
      Moreover, the coupling constant in QCD is inversely proportional to the density: $\alpha_S \sim 1/\rho.$
      Hence  matter with $\rho > \rho_0$ and $T=0$ yields automatically $\alpha_S\ll 1.$ This
      means that the interaction strength is at lowest  and therefore the quarks must be moving freely, practically unaffected by whatsoever external forces.
\end{enumerate}

Consequently, as GQPs inside neutrons are incompressible, we expect nuclear fluid
inside SBs to be indifferent. This implies that, when the distance between neutrons
becomes critically small, e.g.  ${\ell_n}/{\ell_q} = \calO(1),$ then the pions, $\pi^0$, the carriers of the residual strong force between neutrons, become sufficiently energetic and
overcome the repulsive barrier, where they go through a "gluonization" process, which subsequently enables them to mediate the strong force between quarks inside SBs efficiently (Fig. 3).\\

\begin{figure}
\centering {
\includegraphics*[angle=-0, width=6.5cm]{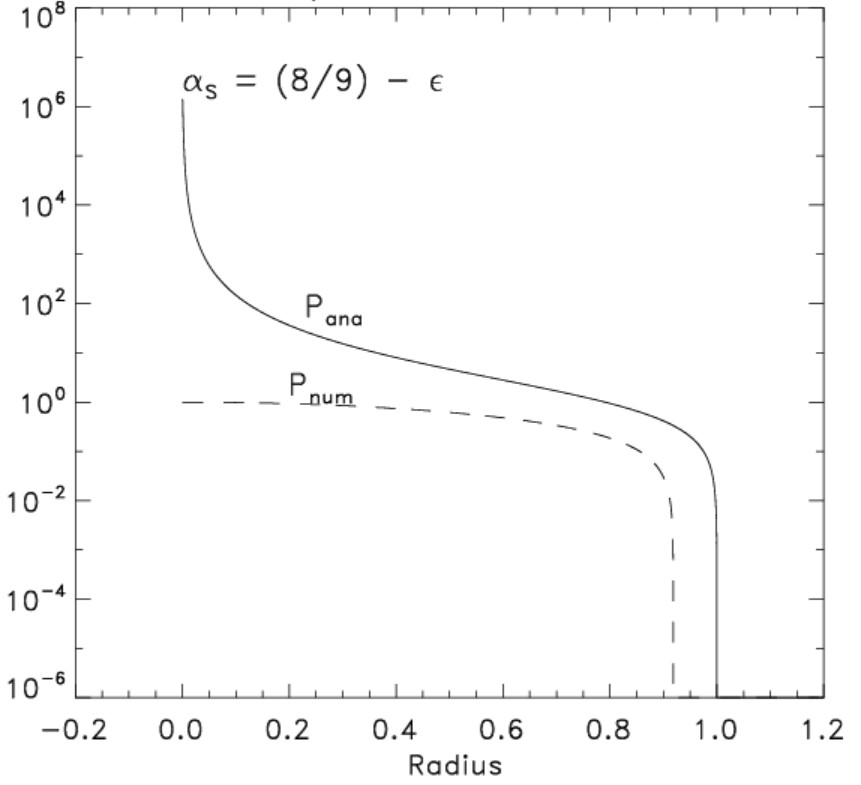}\\
}
\caption{\small The analytical solution of the TOV equation $(P_{ana})$, assuming the fluid to be incompressible (i.e., the internal energy density $\calE_{in}= const.$) for a compactness parameter $\alpha_s (\doteq R_S/R_\star)$ near the critical value 8/9. Obviously, the solution diverges as the center is approached, where the matter becomes ultrabaric. The second profile ($P_{num}$) corresponds to a numerical integration of the TOV equation using $\calE_{in}= const.$) and
assuming $P(r=0) =\calE_{in}.$  }  \label{AnaIncomp}
\end{figure}
 Unlike EOSs in compressible normal plasmas,  classical EOSs in incompressible superfluids are non-local. In the latter case,
            constructing a communicator that merely depends on local exchange of information generally  would not be sufficient for efficiently coupling
            different/remote parts of  the fluid in a physically consistent manner.  A relevant example is the solution of the TOV-equation for classical
            incompressible fluids $(\calE =const.)$.
            In this case, the pressure depends, not only on the global compactness of the object, but it becomes even acausal whenever the global compactness is enhanced
            (see Fig. 2).\\
            This is similar to the case when solving the incompressible Navier-Stokes equations, where an additional Laplacian for  describing the spatial variation             of a non-local scalar field  is constructed to generate  a pseudo-pressure (; actually a Lagrangian multiplier), which, again, does not respect causality \cite[][]{Hujeirat2009}.
\begin{figure}
\centering {
\includegraphics*[angle=-0, width=6.5cm]{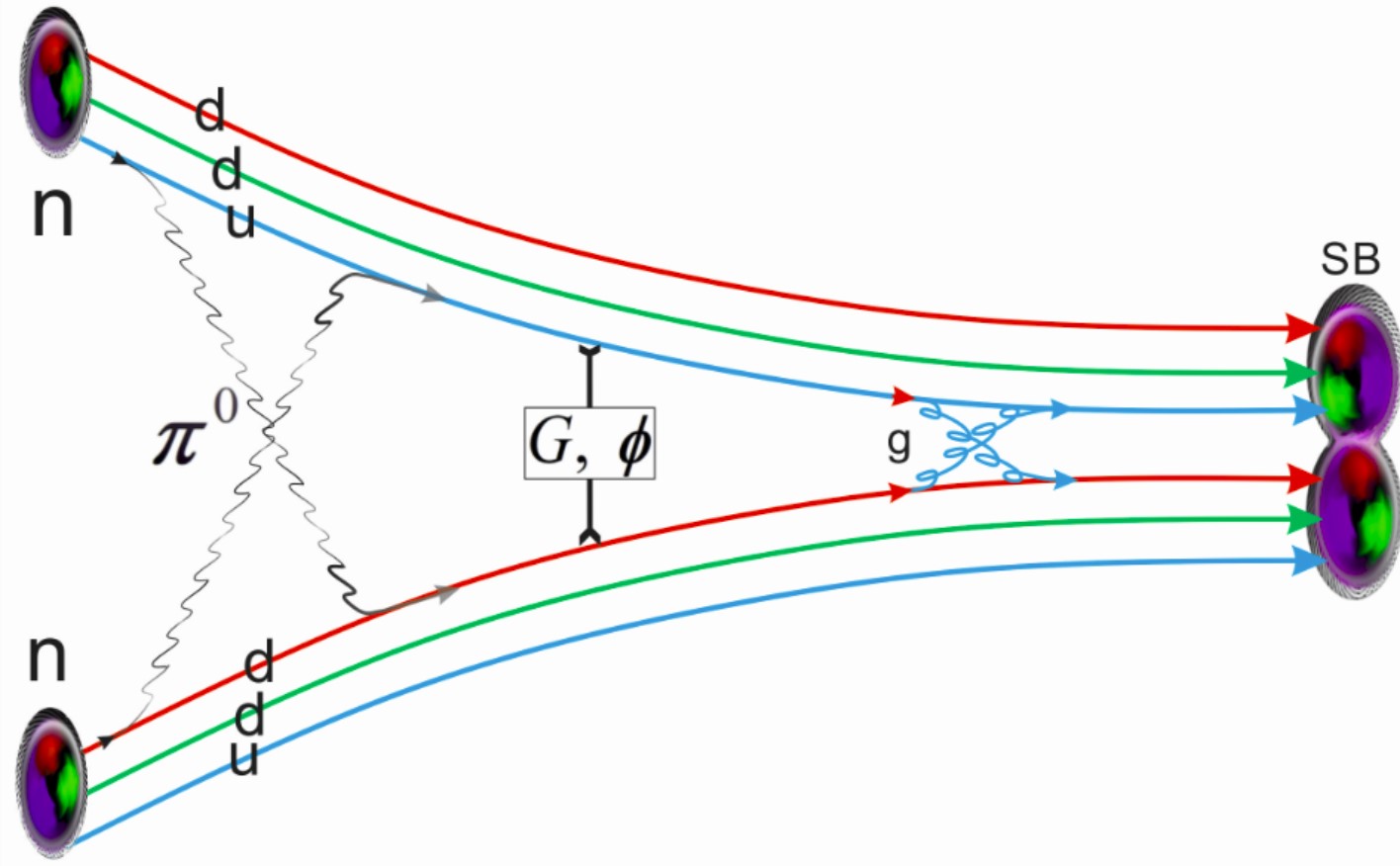}\\
}
\caption{\small A Feynman-like diagram of two interacting neutrons at the center
of a pulsar (time runs from left-to-right). These two neutrons, that are made of two "d" and "u" quarks approach each other under the effect of the gravitational and the scalar fields. Once the distance between them falls below a certain critical value, the residual strong force undergo a "gluonization" procedure necessary for efficiently mediating the strong force between quarks. Here the two particles merge together and form a super-baryon. }  \label{FeynmanDiagramm}
\end{figure}

\section{The onset of superfluidity}
In the presence of magnetic torque and in the absence of energy generation at the
center of pulsars, all types of energies, including magnetic, rotational and rest of thermal energies must diffuse outwards. This energy loss enforces the degenerate core to cool down to $T=0$ and to increase its density. In this case the corresponding de Broglie wave length,
($\lambda_{De} \propto 1/T$), which coincides with the radius of the core, increases  and the enclosed particles start occupying the same quantum state, hence entering the superfluid phase with vanishing entropy: $(S = k_B ln(\Omega)=0)$. In fact this is the lowest energy state possible in the system, which is termed here $L-State.$ \\

Indeed, recent RHIC and LHC-experiments reveal that GQPs inside smashed protons, which are
generally at much higher energy states than the states considered here, were found to be nearly perfect with even smaller viscosity over entropy ratio \citep{Shuryak2017}.  \\

Consequently, we predict the interiors of SBs, which result from fusion of individual neutrons in the $L-State,$ to be in a purely superfluid and superconducting phase.\\

The injected dark energy by $\phi$ may affect the central nuclear fluid in two way:
it first strengthens the curvature of the embedding spacetime, enhancing compression
of the central fluid, and secondly, it provokes the "gluonization" procedure:
\beq
            \Delta E_\phi + \Delta E_G + \pi^0 \rightarrow \Delta E^{SB}_g,
\eeq
where $\Delta E_\phi, \Delta E_G, \pi^0, \Delta E^{SB}_g$ denote the local energy enhancement by $\phi$ and "G" ($\doteq$ gravitational force), the pion energy and the resulting excess of energy gained by the SB, respectively. This reaction is assumed to proceed directly and silently without destroying the superfluidity character of  GQ-fluid inside SBs. \\

On the other hand, the rotational torque of the dissipative ambient medium exerted
on the SBs would enforce them co-rotate uniformly. However, SBs are quantum identities
and therefore must obey the laws of quantum mechanics. This means that their dimensions and energies can accept discrete values only. It turns out that SBs evolve in a discrete manner as prescribed by Onsager-Feymann equation in superfluids. SBs here must eject a certain number of vortex lines in order to increase their dimensions \cite{Hujeirat2017Glitch}. Following this scenario, the internal structure of pulsars and UCOs looks  as follow:\\
(a) Pulsars are born with SB-embryos at their centers, which are made of an incompressible superconducting gluon-quark superfluid, \\
(b) a dissipative neutron fluid  that surrounds the SB,and \\
(c) a geometrically thin boundary layer in-between(BL), where the residual of the strong nuclear force becomes dominant over the viscous forces. The viscosity of the neutron fluid here decreases strongly inwards and vanishes at the boundary of the SB.
 The BL here is practically the zone, where the neutron fluid is prepared to match the physical conditions governing GQ-fluids inside SBs prior to their merger with other neutrons. It turns out that these merger events are identical to the sudden glitch phenomena observed in pulsars and young neutron star systems
 \cite{Hujeirat2017Glitch}. Accordingly, once the rotational frequency of the ambient medium falls below a certain critical frequency $ \Omega_c$, the SB then undergo a sudden spin-down to the next  quantum-allowable frequency, thereby expelling a certain number of vortices, which are then absorbed
 by the ambient medium and causes the observed spin-up of  crusts of pulsars (see Fig. 4). However,
 the excess of rotational energy is then viscously-redistributed into the entire ambient dissipative medium: a process that may last for weeks or even months.\\
 Indeed,  very recent observations appear to confirm our scenario.
 \cite{Eya2017} found that the mean fractional moment of inertia in the glitching pulsars
 correlates weakly with the pulsar spin, implying therefore that glitches are
provoked by a central core of different physical properties. Also \cite{Serim2017}
discovered for the first time that the short X-ray outburst 25 days prior to the glitch
in the accretion-powered pulsar SXP 1062 did not alter the spin-down of the source, which
indicates that glitches are triggered by the core of the NS rather than by the outer shells.
\begin{figure}
\centering {
\includegraphics*[angle=-0, width=6.5cm]{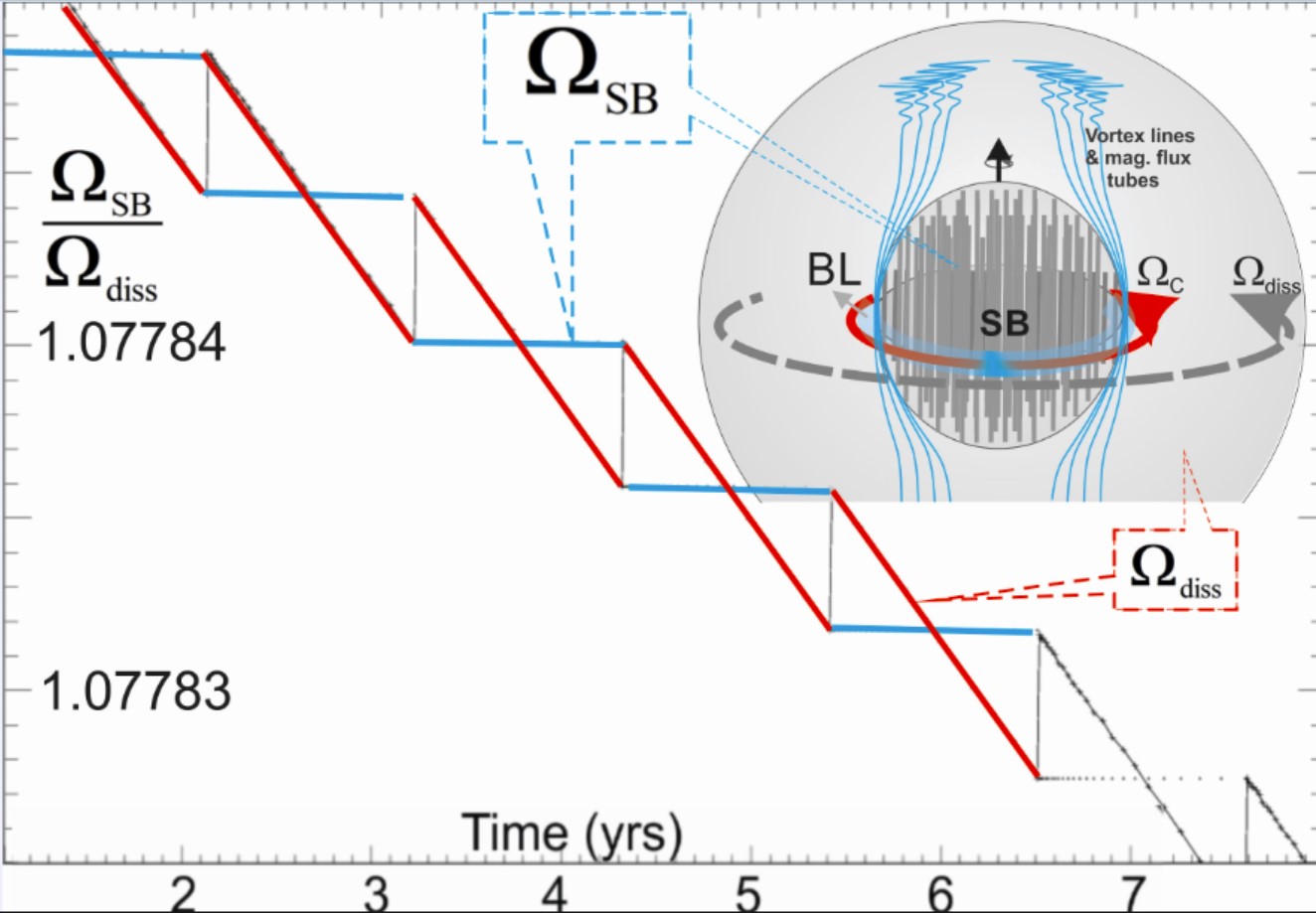}\\
}
\caption{\small  The discrete time-evolution of the rotational frequency of an SB $(\Omega_{SB}/blue~colour)$ and of the ambient dissipative medium $(\Omega_{diss}/red~ colour).$ The rotational frequency and the dimension of the SB evolve as dictated by the Onsager-Feynman equation for  superfluids (Hujeirat 2017c). }  \label{Glitches}
\end{figure}
\section{Crossover phase transition}
 For $ \rho > \rho_0$,  short-range interactions between particles mediated by the exchange of
 vector mesons most likely would enhance the
 convergence of the EOSs towards  $P \rightarrow \calE_b\sim n_b^2$  \cite[see][and the references therein]{Haensel2007,Camenzind2007}.
  The chemical potential here:  $\mu (\doteq (\calE_b +p)/n_b)$ increases linearly with the number density  $n_b.$ Matter with $d\, \mu /d\, n > 0$ is classified here as H-State and depicted in red-color  in Fig. (5). \\
 However In the absence of energy generation, degenerate matter cannot hold a correlation
 of the type $\mu\sim n_b$  indefinitely and it must terminate at a certain critical density $n_{cr}$. This agrees with the two facts: (i) central densities in NSs, $\rho_c,$  increase with their masses and (ii)  $\rho_c$ must be upper-bounded by $\rho_c \leq 12,5 \times \rho_0$ in order to fit the observed mass function \cite[see][and the references therein]{Lattimer2011}.

 On the other hand, in an ever expanding universe, the eternal-state of matter should be the one at which the internal energy attains a
 global minimum in  spacetime (zero-temperature, zero-entropy and where  Gibbs energy per baryon is lowest, i.e. in the L-State).
 Taking into account that $\mu(r) e^{\calV(r)} = const.$ inside massive NSs together with the a posteriori results $e^\calV \ll 1$ (see Fig. 9),
  we conclude that  $\mu \sim \calE_b/n_b = const.$  Under these conditions the gradient of the
  local pressure $P_{local}$ vanishes   and a non-local pressure $P_{NL}$ is necessary to avoid  self-collapse of the object into a BH.\\
 \emph{\footnote{Classically  incompressible fluids with $\calE= const.$ have negative local pressure. Therefore an acausal non-local   pressure is usually used for stabilizing the fluid configurations modelled by the incompressible Navier-Stokes equations.}}
 The slop of the chemical potential $d\mu/dn$  between the H and L-states in the n-space
 could attain positive, negative or even discontinuous values.\\
 However the case with  $d\mu/dn > 0$  should be excluded, as it implies that the eternal state of matter would be more energetic than the H-State, which is a contradiction by construction.
 Similarly, the case $d\mu/dn < 0$  is forbidden as it would violate energy conservation
  (; $d\mu/dn <0 \Leftrightarrow dP/dn <0 \Leftrightarrow$ adding more particles yields a smaller pressure).
  Moreover, let us  re-write the TOV equation in terms of $\mu$:
  \beq
  \DD{d\mu}{dr} = -\DD{G}{c^2 r^2}(\DD{d\calE_b}{d\mu})(\DD{d\mu}{dn})(\DD{m+4\pi r^3P}{1- {r_s}/{r}}).
  \eeq
Obviously,   as $\mu >0,$ a negative $d\mu/dn $ would destabilize  the hydrostatic equilibrium, unless external sources are included, e.g. bag energy and/or  external fields.\\
 Therefore, although  a first order phase transition may not be completely excluded, a crossover phase transition
 into an incompressible superfluid phase with $\mu=\mu(n=n_{cr}) = const.$  would be more likely.
 Here, $\mu$ and $P$ on both sides of the transition front  are equal and, with the help of an
   external field, both $(\calE_{tot}/n)^+ $ and $(\calE_{tot}/n)^-$ across the front  can be made continuous (Fig. 2)).\\

 In the present study, the crossover phase transition corresponds to "silent" mergers of
 baryons forming SBs. Inside SBs, the fluids have $\mu = n= const.$ and
 $T=0$ and therefore vanishing entropy. To overcome the repulsive barrier between
 individual neutrons and provoke their merger, external forces are required. In the present study, we assume that, in addition to compression by self-gravity, there is a universal scalar field, $\phi,$
 at the background of supranuclear dense matter, which provide the energy required for
 forming an eternal stable GQ-cloud inside SBs. In this case, the dark energy potential
 associated with $\phi$ should have similar effects as the generalized quark-potentials inside individual neutrons, namely:
 \beq
   \calV_\phi =  a_\phi r^\gamma + b_\phi + \DD{c^{EM}_\phi}{r} ,
 \eeq
  where $a_\phi, b_\phi, c^{EM}_\phi$ are constant coefficients \citep[see][and the references therein]{Sakumichi2015}.  Without loss of generality,
   we may set $c^{EM}_\phi=0$ and $\gamma=2.$ As the scalar field is universal, its special
    and temporal variations are assumed to vanish. Therefore, the expression
    $ \calE_\phi = \half \dot{\phi}^2 + V(\phi) + \DD{1}{6} (\nabla\phi)^2$ reduces
    to $\calE_\phi =V(\phi)= \calV_\phi.$ By requiring the chemical potential at $r=0$ to be upper-limited by the deconfinement energy of quarks inside a single neutron, then
   a Gibbs function of the following type can be constructed:

 \beq
       f(n) = \DD{\calE_b + \calE_\phi}{n} - 0.939\mbox{ GeV}.
 \eeq
Using the scalings $[\rho]=10^{15}g/cm^3~(\doteq 0.597/fm^3),$ chemical potential (energy per particle) $[\mu] = 1\,GeV,$  we then obtain
  $[a_\infty]=1.674~GeV\, fm^3$ and  $ [a_\phi]= 5.97\times 10^{-39}~GeV/fm^5$ and $[b_\phi] = 0.597~GeV/fm^3.$\\
\begin{figure}
\centering {
\includegraphics*[angle=-0, width=8.15cm]{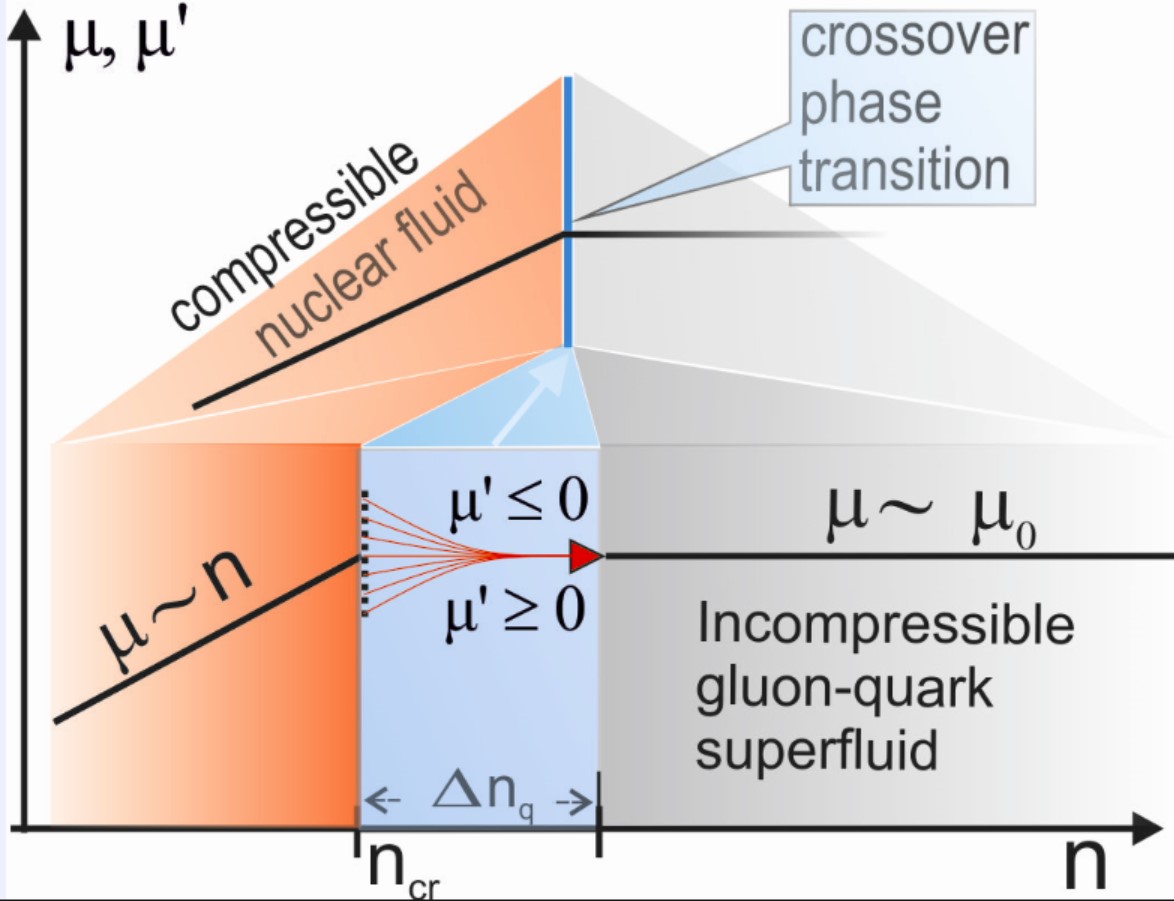}\\
}
\caption{\small  A schematic description of  the chemical potential $\mu,$ $\mu^{'}(\doteq \partial \mu/\partial n)$ versus  the number density $n$ at
the centers of NSs.
 At a critical central density, $n_{cr},$ the universal scalar field $\phi$ is set to provoke a crossover phase transition
 from compressible and dissipative neutron fluid into incompressible GQ-superfluid. The transition front creeps from inside-to-outside
 to reach the surface of the object on the scale of one Gyr.}  \label{PhaseTransition}
\end{figure}

Substituting $\calE_b = a_0 n^2$ and $V_\phi(r)$ in Eq. (5) at $r=0$,  then $f(n)$ reduces to:
 \beq
       f_(n) =  a_\infty\,n+ \DD{b_\phi}{n} - 0.9396.
 \eeq
  The Gibbs function here may accept several minima at $n_{min}= (b_\phi/a_\infty)^{1/2},$  though $f(n=n_{min})$ doesn't necessary vanish.
  However $f(n)> 0$ and $f(n)<0$ should be excluded, as they are energetically not suited for a smooth crossover phase transitions. \\
  On the other hand, by varying $a_\infty\mbox{ and }b_\phi,$ a set of minima could be found.
  One way to constrain $b_\phi$ is to relate it to the canonical energy scale characterizing the effective coupling of quarks, i.e.
  $b_\phi=0.221$ \cite[see][and the references therein]{Bethke2007}. Indeed,  as shown in Fig (3),   $f(n)$ attains a zero-minimum at $n \approx 3~ n_0$
  for $a_\infty=1.0.$  \\

\begin{figure}
\centering {
\includegraphics*[angle=-0, width=8.15cm]{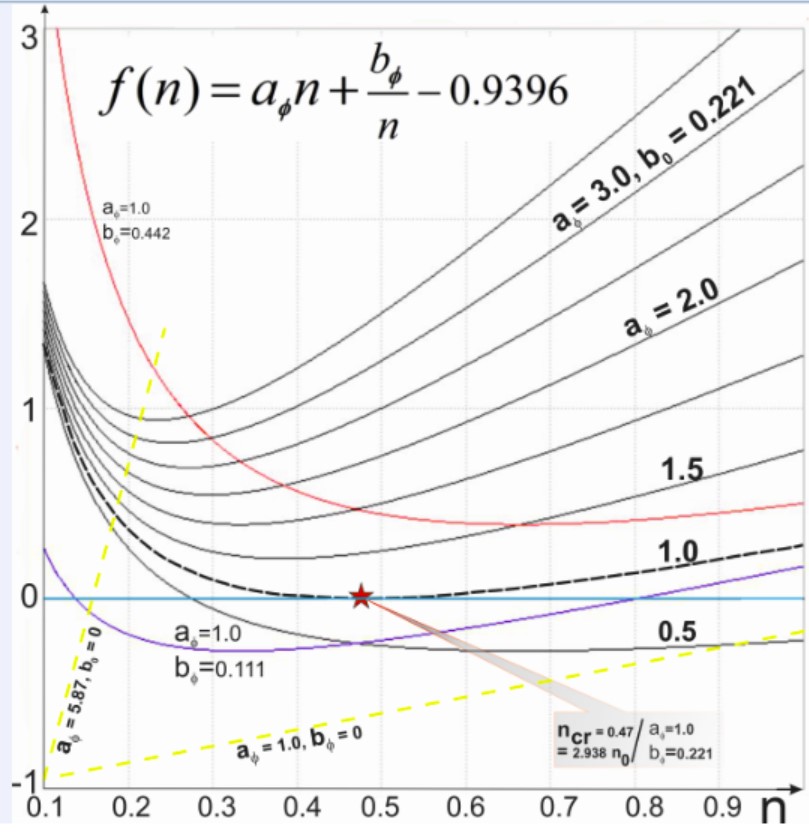}\\
}
\caption{\small  The modified Gibbs function $f(n)$ versus  baryonic number density n (in units of $n_0$) is shown for various values of $a_\infty$ and $b_\phi.$
Obviously, $a_\infty= 1 $ and $b_\phi=0.221$ appear to be the most appropriate parameters compatible  with QCD. The value $b_\phi=0.221$ corresponds
 to the canonical energy scale characterizing the effective coupling of quarks inside individual hadrons. The Gibbs function $f(n)$ here attains
  a zero-minimum at $n=2.938\, n_0,$  at which  $\phi$ is set to provoke a phase transition into the SuSu-fluid state. }  \label{GibbsFunction}
\end{figure}

\begin{figure}
\centering {
\includegraphics*[angle=-0, width=7.95cm]{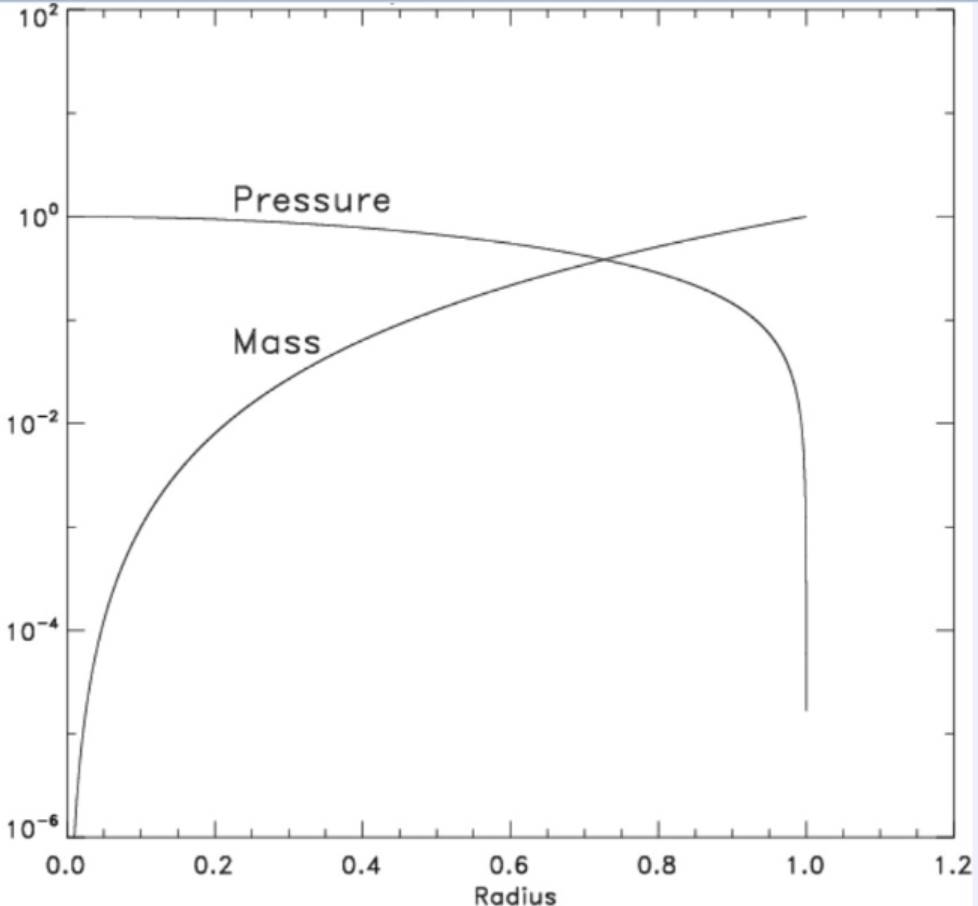}\\
\includegraphics*[angle=-0, width=7.95cm]{Bild71}
}
\caption{\small For test purposes the profiles of the pressure and enclosed mass of a NS that have been obtained by solving the TOV equation, using a polytropic EOS $(P= \calK \rho^\gamma).$ In the lower figure the
the numerical errors for two different numerical integration methods are shown: the first order Euler and the fourth order Runga-Kutta methods using $10^4$ grid points compared to a reference solution that has been obtained using $10^7$ grid points. While the accuracy of  Ruga-Kutte method is highly superior over Eulers, the effect
of errors on the solutions presented here can be safely neglected.
  }  \label{P_M_NS_Test}
\end{figure}

 \section{The governing equations and the solution procedure}
 The investigation here is based on numerical solving the TOV equation modified to include scalar fields $(\phi).$  The modified stress energy tensor  reads:
\beq
 T^{mod}_{\mu\nu} = T^0_{\mu\nu} + T^\phi_{\mu\nu}.
\eeq
The superscripts "0" and "$\phi$" correspond to baryonic and scalar field tensors:
\[
T^0_{\mu\nu} = - P^0 g_{\mu\nu} + (P^0 + \calE^0)U_\mu U_\nu \textrm{ ~~and ~~ }
\]
\beq
T^\phi_{\mu\nu} = (\D_\mu \phi)(\D_\nu \phi) -g_{\mu\nu}[\half (\D_\sigma \phi)(\D^\sigma \phi)-V(\phi)].
\eeq
$U_\mu$ here is the 4-velocity,  the subindices $\mu,~\nu \textrm{ ~  run from 0 to 3}$  and
$ g_{\mu\nu } $ is a background metric of the form:
\beq
 g_{\mu\nu} = e^{2\calV}dt^2 - e^{2\lambda}dr^2 - r^2 d\theta^2 - r^2 sin^2\theta d^2\varphi^2,
\eeq
where $\calV,~\lambda$ are functions of the radius.\\
Assuming the configuration to be in hydrostatic equilibrium, then  the GR field equations, $G_{\mu\nu} = - 8\pi G \,T_{\mu\nu}$  reduce into  the generalized TOV equations:
\beq
\DD{dP}{dr} = - \DD{G}{c^4 r^2} [\calE + P]{[ m(r) + 4\pi r^3 P]}/{(1 - r_s/r)},
\eeq
where ${m(r)} = 4\pi \int \calE r^2 \,dr $  is the total enclosed mass:
$\calE = \calE^0 + \calE^\phi,~ P = P^0 + P^ \phi, $  where $\calE^\phi = \half \dot{\phi}^2 + V(\phi) + \half (\nabla\phi)^2,$
$ P^\phi = \half \dot{\phi}^2 - V(\phi) - \DD{1}{6} (\nabla\phi)^2.$\\
$V(\phi)$ here denotes the interaction potential of the scalar field with the baryonic matter, i.e., the rate at which dark energy
is injected into the system and $\dot{\phi}$ is the time-derivative of $\phi.$ \\

\begin{figure}
\centering {\hspace*{-0.75cm}
\includegraphics*[angle=-0, width=8.65cm]{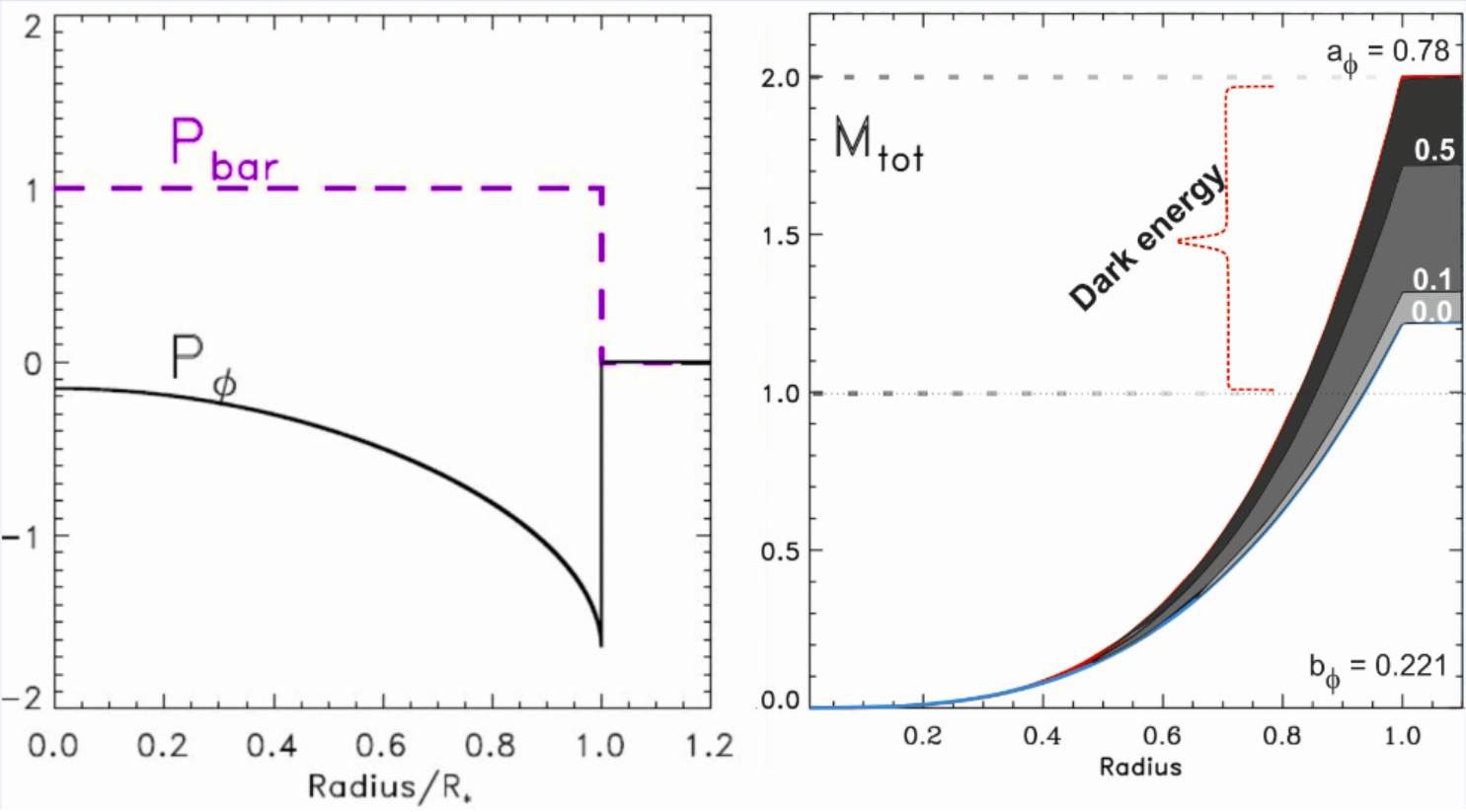}
}
\caption{\small  The radial distributions of the baryonic pressure ($P_{bar}$) and negative pressure  ($P_\phi$)  inside
an incompressible GQ-superfluid core (left). The enclosed mass of the baryonic matter and the gradual mass-enhancement
due to dark energy  is shown for different values of  $a_\phi$ (right).
}  \label{Pressure-Mass-DEOs}
\end{figure}
Our reference object is a NS with $1.44~\calM_\odot,$ a radius $R_\star = 2\times R_S,$  where $R_S$ is the Schwarzschild radius. The universal scalar field, $\phi$ is assumed to be spatially and temporarily constant, whereas $V_\phi$ is taken from Eq.(4). The non-dimensional value of $a_\phi$ is  chosen to fulfill the
a posteriori requirement: $R_\star= R_S + \epsilon,$ where $0\leq \epsilon \ll 1.$ In most of the cases considered here,  $b_\phi$ is set to
be identical to the canonical energy scale at which momentum transfer between quarks saturates, i.e.,  $b_\phi=0.221$ GeV.
The  fluid in the post transition phase is governed by the EOS: $P^0 =P_L= \calE^0 = \rho_{cr}c^2= const.$\\
For a given central density, the  solution procedure adopted here is based  on integrating the TOV-equations for the pressure, enclosed mass and for the pseudo-gravitational potential from inside-to-outside,  using either the first order Euler or the fourth order Runge-Kutte integration methods.\\
For verification purposes the TOV equation has been solved using a polytropic EOS as well
as for incompressible case $(\calE_b = const.),$ using both integration methods (Fig. 7).\\

Similarly, in Figures (8) and (9)/top we show the profiles of the negative pressure, the enclosed effective mass (i.e. the energy due to baryon and energy enhancement by the scalar field $\phi$) and the inverse of the metric coefficient $g_{rr}$ for different values of the dark energy coefficient $a_\phi$ (see Eq. 4).\\
Obviously, assuming the object to have $\alpha_S = 1/2$ initially, its final compactness
must then increase with increasing $a_\phi,$ i.e. with increasing the rate of dark energy injection. Using the limiting values $a_\phi=0.78$ and $b_\phi=0.22,$ the scalar field is
capable of injecting the energy required to deeply sink the object into spacetime and turn it invisible, where $\alpha^{cr}_S = 1 - \epsilon, $ hence attaining its maximum possible value.
Note that $\epsilon$ must be extremely small, but still finite in order to prevent the collapse of the object into  a BH.\\
These values of $a_\phi$ and $b_\phi$ yield a final effective mass that
is twice as the initial baryonic mass.\\
Once the object has completely metamorphosed into a stellar-size SB and its
compactness attained the critical limit $\alpha^{cr}_S$, then the spatial variation of the metric coefficient $g_{rr}$ across its surface becomes nearly singular (Fig. 9), which
implies that all sorts  of surface radiation will be extraordinary redshifted, hence
the object becomes observationally indifferent from its BH-counterpart.  On the other hand,
the metric coefficient $g_{tt},$ which is a monotonic function of the gravitational potential, appears to be fairly flat inside the object, but it undergoes a dramatic change across the surface to finally attains its weak-field values outside the object.

In addition, we have solved the TOV equation using two different EOSs: an EOS that corresponds to an incompressible GQ-superfluid inside a gradually growing SB together with a polytropic EOS for the ambient medium at different evolutionary epochs (Fig. 10). While the scalar field here is set to inject  energy into the SB and therefore increasing its effective mass, the surrounding medium appears to stably and comfortably adjust to the new condition of the SB.

In Fig. (11) we display the Mass of fully-developed SuSu-objects and their progenitors
versus $\rho_{cr}$ superimposed on the locations of NSs as reported by \cite{Lattimer2011}. Most remarkable here is that Hulse-Taylor type pulsars are able to form
SBs at their centers at much lower central density than usually required for a phase
transition into quark fluids, whilst still end up twice as massive as their initial mass
(Fig. 11).
\begin{figure}
\centering {\hspace*{-0.75cm}
\includegraphics*[angle=-0, width=8.15cm]{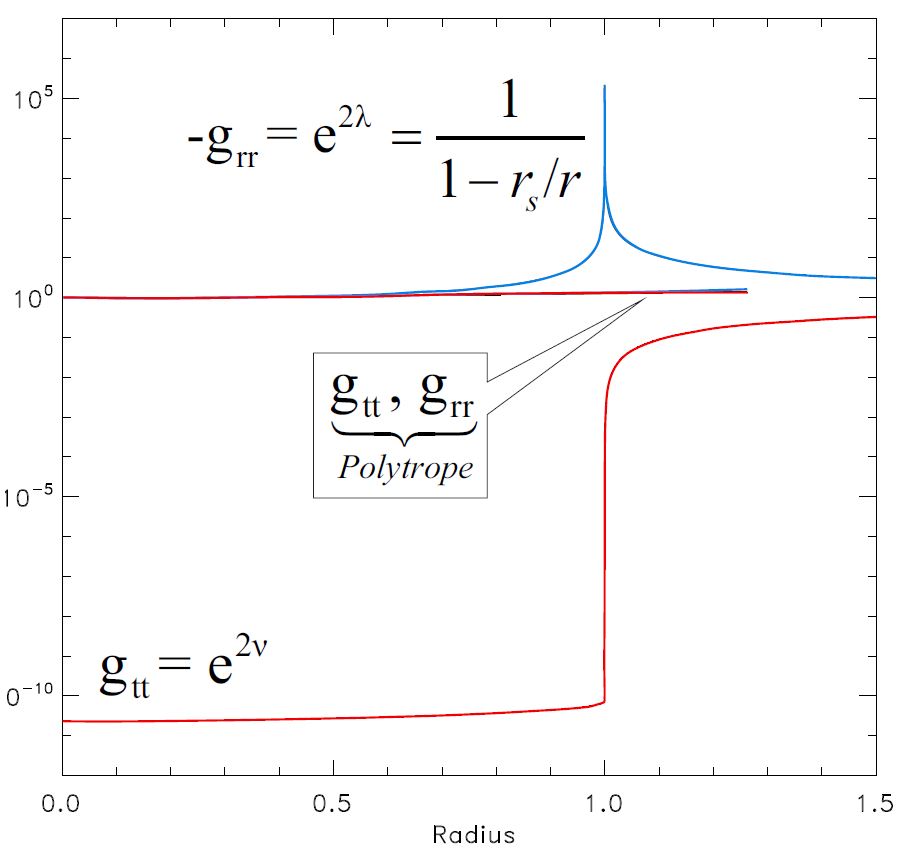}
\includegraphics*[angle=-0, width=8.15cm]{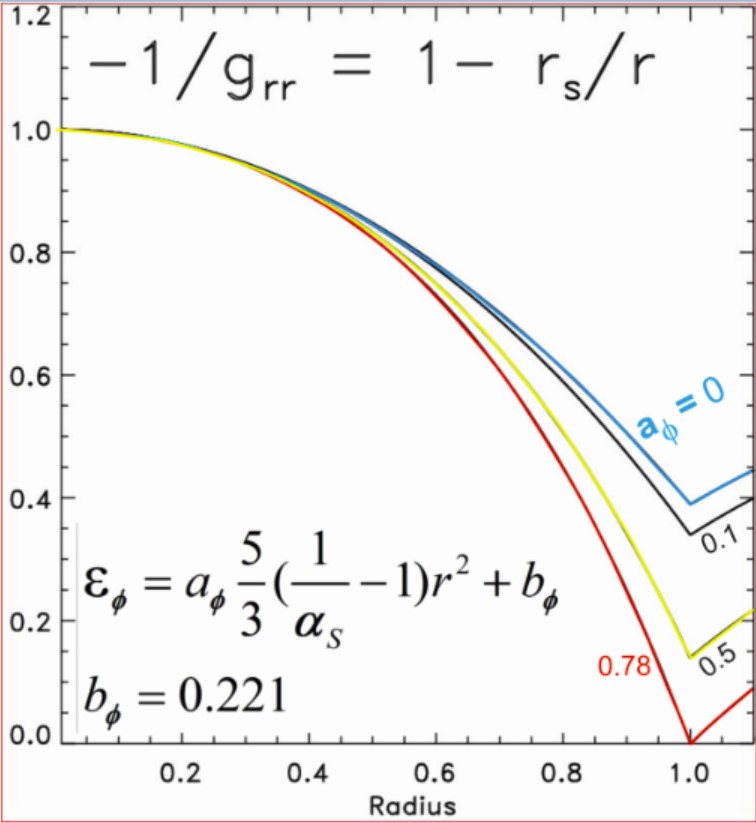}
}
\caption{\small In the top panel we show the radial distributions of the metric coefficients $g_{rr}$ and $g_{tt}$
inside a normal NS (; $P_L=\calK \rho^\gamma$ and $P^\phi =0$) and inside an SuSu-object (; $P_{local}= const.$ and $P^\phi =-V_\phi$).
Obviously,  normal models of NSs  have larger radii and considerably less compact than their SuSu-counterparts, which
can be inferred from the very limited spacial variations of $g_{rr}$ and $ g_{tt}.$   In the lower panel, the compactness of  a typical SuSu-object,
expressed in terms of  -$1/g_{rr}$ is shown for different values of $a_\phi.$ The object turns invisible if  $V_\phi$ is calculated with
$a_\phi=0.78$ and $b_\phi=0.22.$
}  \label{Metric-Coefficient-grr-gtt}
\end{figure}

\begin{figure}
\centering {\hspace*{-0.75cm}
\includegraphics*[angle=-0, width=8.15cm]{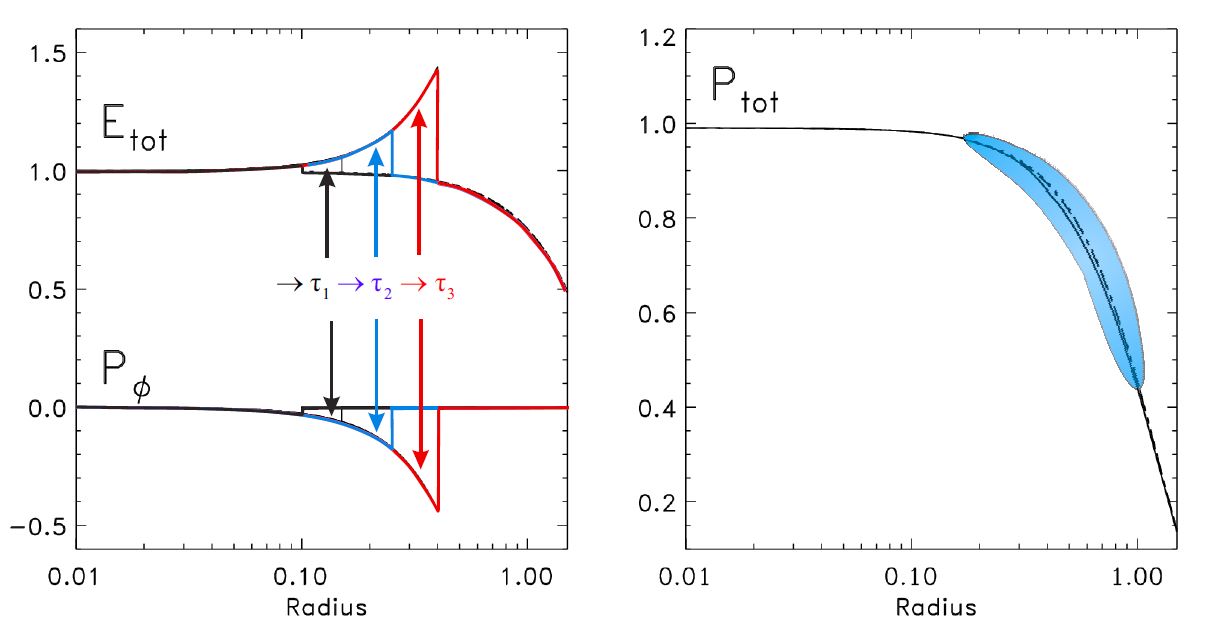}\\
}
\caption{\small The profiles of the total energy density $E_{tot},$ the  pressure $P_\phi$ induced by  $\phi$
and the combined pressure $P_{tot}$ versus radius  are shown for  different evolutionary epochs $ \tau_1 < \tau_2 < \tau_3.$
Inside $r_f$:  $P_{L} =0$ and $P^\phi= -V_\phi,$ whereas outside $r_f$: $P_L=\calK \rho ^\gamma$ and $P_\phi =0.$
 In each epoch, the object has an SB-core  overlayed by a shell of normal compressible  matter obeying a polytropic EOS.
 Obviously, the object appears to stably and comfortably adjust itself to the mass-redistribution inside $r_f,$ where the matter is in incompressible GQ-superfluid state.
}  \label{Creeping-Front}
\end{figure}

\begin{figure}
\centering {
\includegraphics*[angle=-0, width=8.15cm]{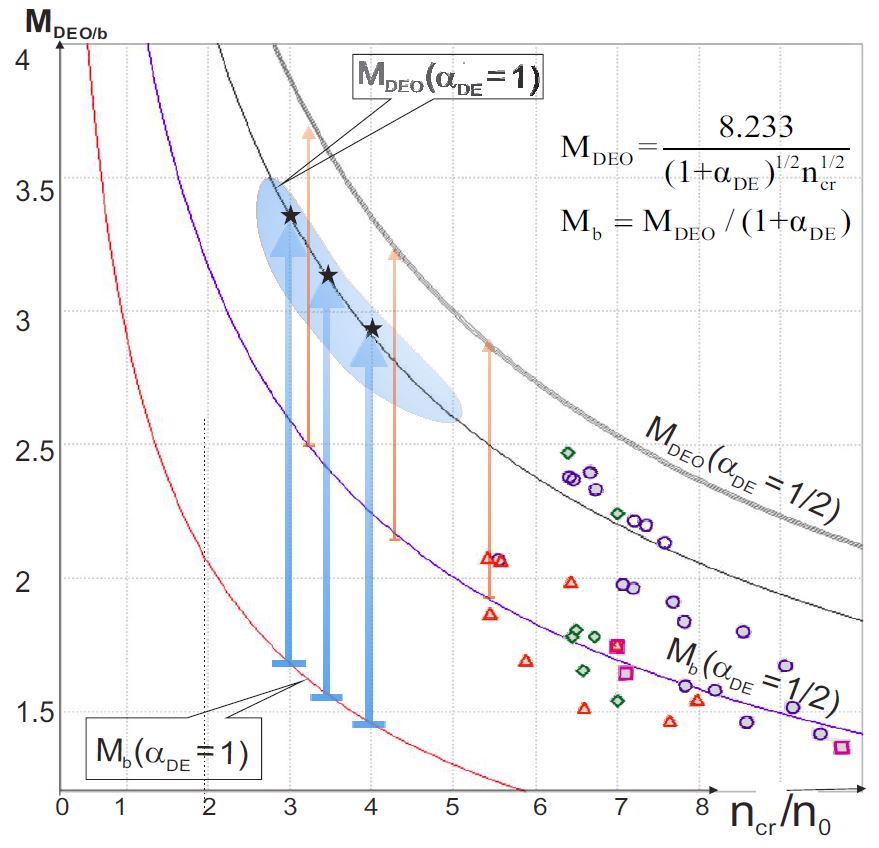}\\
}
\caption{\small   Upper mass limit of SuSu-objects versus critical density $n_{cr}$ (in units of $n_0$) is shown. The $\phi-$baryon interaction is set
 to occur at $n_{cr},$ which in turn  provokes  the phase transition into the
 incompressible GQ-superfluid state. The most probable  mass-regime for SuSu-objects
is marked here in blue colour. Accordingly, the progenitor of a SuSu-object with $3.36\,\MSun$   should be a  pulsar/NS of $1.68\,\MSun,$ provided it has an initial compactness $\alpha_S = 1/(1+\alpha_{DE})= 1/2$ and  $n_{cr} = 3\,n_0.$ Similarly, a Hulse-Taylor type pulsar would end as an SuSu-object of $2.91\,\MSun,$ if  its initial compactness was $\alpha_S = 1/2$  and  if $n_{cr} = 4\,n_0.$
On the other hand, moderate and massive NSs with initial compactness $\alpha_S \geq 2/3,$ i.e., $ \alpha_{DE} \leq 1/3,$ need less dark energy to become invisible SuSu-objects, though  an unreasonably high $n_{cr}$ is required for triggering  $\phi-$matter interaction. NSs  falling in this category are to be compared
with  the colored small cycles and triangles, which show the approximate locations of various NS-models as depicted in Fig.  (4) of Lattimer \& Prakash (2011).
}  \label{NSs-Migeration}
\end{figure}
\begin{figure}
\centering {\hspace*{-0.75cm}
\includegraphics*[angle=-0, width=6.15cm]{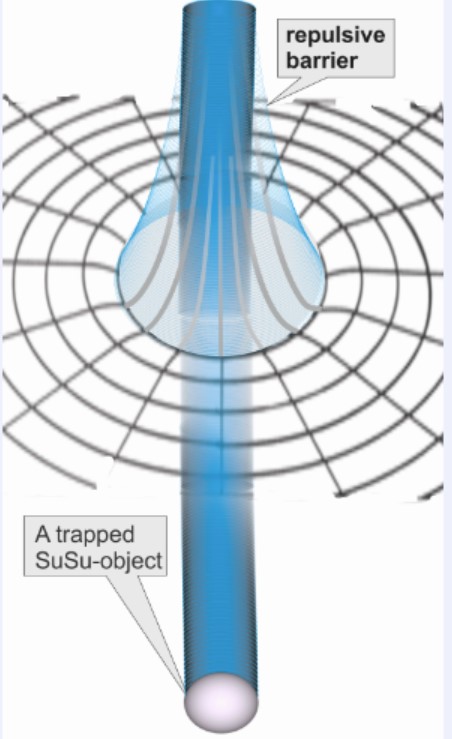}\\
}
\caption{\small A schematic description of an invisible SuSu-object trapped in spacetime and surrounded by a repulsive barrier that protects/confines the enclosed sea of GQ-superfluid. The spacetime inside SuSu-objects is fairly flat, but  extraordinary curved and nearly
singular across their surfaces.
}  \label{TrappedSuSu}
\end{figure}
 \section{Summary \& discussions}
 The here-presented model is motivated by  the following issues:
   \bit
  \item  Why  neither NSs nor BHs have ever been observed in the mass-range 2 - 5$\,\MSun.$
  \item  Most sophisticated EOS used to model the internal structure of NSs are based on central densities that are far beyond the
            nuclear density: an unknown density regime with great uncertainty.
    \item What is the origin of glitches observed in pulsars and young neutron stars and
  wether these carry information that may disclose the internal structure of their cores?
  \item Could massive NSs end as maximally compact dark objects, i.e. as BH-candidates?
  \item How does the state of matter in pulsars and NSs evolve on the cosmological time scale
        and whether they have hidden connection to dark matter and dark energy in cosmology?
  \eit

  In this respect a scenario has been presented, which can be summarized as follows:
 \ben

 \item  Pulsars are born with embryos at their centers (here termed super-baryons -SBs).
  The interiors of SBs are made of GQ-superfluids and governed by the EOS $P=\calE = a_\infty\,n^2:$ a purely incompressible fluid state.\\
  As a consequence, there is a universal maximum density
  $\rho^{uni}_q (= \calO(\rho_{cr})),$
   where the momentum transfer between  quarks saturates, the coupling constant $\alpha_{asym}$ attains its universal minimum, where quarks are moving freely.\\
    Recalling that the  spatial variation of  the coefficient $g_{rr}$ of the Schwarzschild metric on the nuclear length scales is negligibly small
  ($dg_{rr}/dl \ll 10^{-19}$),  the GQ-superfields may not accept stratification by gravitational fields.\\

\item  In the presence of a universal scalar field $\phi$ at the background of supranuclear densities, the injected dark energy by $\phi$ is capable of  provoking
   a phase transition from compressible dissipative neutron fluids into
   incompressible GQ-superfluids. The effect of both the gravitational and scalar fields  is to mainly enhance and convert the residual strong force between nucleons into a strong force that holds quarks together. This action is termed here as a "gluonization" procedure, which is equivalent to energy injection into the system, thereby maximally enhancing the effective mass of the object and turns it invisible.\\
   It turns out that using an interaction potential of the type $V_\phi = a_\phi r^2 + b_\phi$ appear to be most appropriate for maximizing the compactness of the object without
    significantly changing its dimensions (Fig. 5).
\item In a recent study, I have shown that the glitch phenomena observed to associate the evolution of pulsars and NSs are in perfect-agreement with the formation scenario of embryonic SBs at their centers.
\item  Once the entire object has metamorphosed  into a stellar-size SB,  the spacetime in its interior would be fairly flat, but become exceedingly curved across its surface (Fig. 5).
    Hence SBs are practically trapped in spacetime, extraordinary redshifted and therefore
    completely invisible.

\item According to the here-presented scenario, all visible pulsars and NSs must contain SBs. The gravitational significance of these SBs depends strongly on their evolutionary phase and in particular on their  ages and initial compactness.
Accordingly, pulsars and young NSs should be less massive than old ones, and  the very old NSs should  turn invisible by now.\\
To quantify the mass-enhancement by $\phi,$ let $M_b$ be the mass of the NS at its birth
 and $ M_\phi $ being the mass enhancement due to $\phi$. Requiring $R_\star > R_S,$ then
the following inequality holds:
\beq
(1 + \alpha_{DE}) \leq (\DD{3 \rho_{cr}}{32 \pi })^{1/3} \DD{c^2}{G M^{2/3}_b},
\eeq
or equivalently,
\beq
1\leq \DD{E_{tot}}{E_b}\leq   2.06 \DD{\rho^{1/3}_{15}}{M^{2/3}_{b/1.44}},
\eeq
 where $\alpha_{DE} \doteq M_\phi/M_b.$  $E_{tot},~\rho_{15},~M_{1.44}$ denote the total energy, the density in units of $10^{15}$ g/cc
 and the baryonic mass of the NS in units of   $1.44\,M_\odot,$ respectively.\\
  Assuming NSs to be born with ${E_{tot}}= {E_b},$  then by interacting with $\phi,$  they become more massive and more compact to finally  reach
  $R_\star = R_S + \epsilon$   at the end of their luminous phase $(\epsilon\ll 1)$, which would last for approximately $10^9$ yr or even less, depending on their initial
  compactness Fig. 4 and Fig. 7).\\

 \een



 Similar to atomic nuclei,  I conjecture that the enormous surface stress confining the sea of GQ-superfluid inside stellar-size SBs render their surfaces impenetrable by external low energy particles, hence maintaining the eternal stability and invisibility of SuSu-objects.
 This implies that  the core of a fully-developed  SB would be shielded by a protecting
repulsive barrier.
However, due to the deep gravitational wells of SBs,  such incidents would be practically
observable, though they are ruled out by observations completely.
Even if there were no repulsive barriers, we expect SBs to still be stable against mass-enhancement from external sources. Let a certain amount of
baryonic matter, $\delta \calM_b,$   be added to the object via accretion from external sources. Then the relative increase of $R_\star$ compared to
             $R_S$ scales as:  $\DD{\delta R_\star}{\delta R_S} \simeq \DD{\rho_{cr}}{\tilde{\rho}_{new}},$  where
            $\tilde{\rho}_{new}$ is the average density of the newly settled matter. Unless  $\tilde{\rho}_{new} \geq \rho_{cr},$
            which is forbidden under normal astrophysical conditions,  the SuSu-object would react stably. However, in the case of super-Eddington accretion or merger,
            the newly settled  matter  must first decelerate, compressed and subsequently becomes virially hot, giving rise therefore to
             $\tilde{\rho}_{new} \ll  \rho_{cr}.$  On the other hand, such events would lower the confinement stress at  the surface
             and would turn the quantum jump of the energy density  at $R_\star,$ which falls abruptly  from  approximately $\calE \approx10^{36}$ erg/cc at $R_\star$ down to zero outside it,  into an extraordinary steep pressure gradient in  the continuum.
             While such  actions would smooth the strong curvature of spacetime across $R_\star,$  they
              would enable SuSu-objects to eject quark matter into space with ultra-relativistic speeds, which is forbidden.
              Nonetheless, even if this would occur instantly, then the corresponding time scale $\tau_d$ would be
            of order $\Lambda_j/c,$ where $\Lambda_j$ is the jump width in centimeters. Relating $\Lambda_j$ to the average spacing between
            two arbitrary particles $(\sim n^{-1/3}),$ this yields  $\tau_d \approx 10^{-24}\,$s, which is  many orders of magnitude shorter
            than any known thermal relaxation time scale between arbitrary luminous particles.\\

            Although electromagnetic activities and jets have never  been observed in dark matter halos, they are typical events for systems
            containing  black holes. Recalling that supermasive GBECs are dynamically unstable \cite[][]{Hujeirat2012},
             our results here address the following two possibilities:
             \bit
            \item  If the onset of $\phi$-baryon interaction indeed occurs at $n_{cr},$ then the majority of the  first generation of stars must have collapsed into
                UCOs and subsequently became SuSu-objects, rather than collapsing into stellar BHs with $M \leq 5\times \MSun$. These objects are
                expected to conglomerate into dark matter halos over several big bang events and to embed the  galaxies in the observable universe. This conclusion is in line with  recent observations of NASA's Spitzer Space Telescope, which reveal the existence of primordial galaxies, such as GN-z11, whose age might be
                even bigger than  that of our universe  \citep{citeCastellano2016}.\\

            \item The passivity of DM to electromagnetic radiation may indicate that the collective effect of the repulsive forces of a cluster of SuSu-objects is
                repulsive on smaller length scales and attractive on the larger ones.  Hence
                  approaching luminous matter will be forced to deviate from face-to-face collisions with the cluster of SuSu-objects, though n-body and SPH-numerical calculations are needed here to verify this argument.
            \eit

\end{document}